\documentclass[twocolumn]{aastex631}

\usepackage{textgreek, gensymb, needspace}
\usepackage[]{fixltx2e}
\usepackage[bottom]{footmisc}

\received{January 17, 2024}
\revised{March 7, 2024}
\accepted{March 19, 2024}

\submitjournal{ApJ}

\shorttitle{H$\alpha$ Activity in M-M Wide Binaries}
\shortauthors{Pass et al.}

\begin{document}
\widowpenalty=0
\title{The Mass Dependence of H\textalpha\ Emission and Stellar Spindown for Fully Convective M Dwarfs}

\author[0000-0002-1533-9029]{Emily K. Pass}
\affiliation{Center for Astrophysics $\vert$ Harvard \& Smithsonian, 60 Garden Street, Cambridge, MA 02138, USA}

\author[0000-0002-9003-484X]{David Charbonneau}
\affiliation{Center for Astrophysics $\vert$ Harvard \& Smithsonian, 60 Garden Street, Cambridge, MA 02138, USA}

\author[0000-0001-9911-7388]{David W. Latham}
\affiliation{Center for Astrophysics $\vert$ Harvard \& Smithsonian, 60 Garden Street, Cambridge, MA 02138, USA}

\author{Perry Berlind}
\affiliation{Center for Astrophysics $\vert$ Harvard \& Smithsonian, 60 Garden Street, Cambridge, MA 02138, USA}

\author[0000-0002-2830-5661]{Michael L. Calkins}
\affiliation{Center for Astrophysics $\vert$ Harvard \& Smithsonian, 60 Garden Street, Cambridge, MA 02138, USA}

\author[0000-0002-9789-5474]{Gilbert A. Esquerdo}
\affiliation{Center for Astrophysics $\vert$ Harvard \& Smithsonian, 60 Garden Street, Cambridge, MA 02138, USA}

\author[0000-0003-3594-1823]{Jessica Mink}
\affiliation{Center for Astrophysics $\vert$ Harvard \& Smithsonian, 60 Garden Street, Cambridge, MA 02138, USA}



\begin{abstract}
\noindent Fully convective M dwarfs typically remain rapidly rotating and magnetically active for billions of years, followed by an abrupt and mass-dependent transition to slow rotation and quiescence. A robust understanding of this process is complicated by difficulties in estimating M-dwarf ages and potential dependencies on other variables such as birth environment or metallicity. To isolate the effect of mass, we consider M dwarfs in wide binaries. We identify 67 widely separated, fully convective (0.08--0.35M$_\odot$) M-dwarf binary systems using Gaia and measure the H\textalpha\ feature for each component. We classify the pairs into three categories: systems where both components are active, systems where both are inactive, and candidate transition systems, where one component is active and the other inactive. We gather higher-resolution spectra of the candidate transition systems to verify that their behavior does not result from an unresolved third component, yielding one new triple with surprising activity levels. Neglecting this triple, we find 22 active, 36 inactive, and 8 transition pairs. Our results are consistent with the epoch of spindown for these binaries being primarily determined by mass, with mild second-order effects; we place a 1$\sigma$ upper limit of 0.5Gyr or 25\% on the dispersion in the mass-dependent spindown relation. Our findings suggest that the large dispersion in spindown epoch previously observed for field stars of a given mass may stem from differences in birth environment, in addition to modest intrinsic stochasticity. We also see evidence that the wide binary population is dispersed over time due to dynamical processing.
\end{abstract}


\section{Introduction}
\label{sec:intro}
Sun-like stars begin their lives rotating rapidly and gradually spin down over time as they shed angular momentum. The field of gyrochronology -- estimating the ages of stars from their rotation period -- is built upon this principle \citep{Skumanich1972, Barnes2003}. However, fully convective M dwarfs do not behave like Sun-like stars. Rather, field M dwarfs display bimodal rotation periods, with few interlopers located between the modes. This result implies that there is an abrupt transition between short ($<10$ day) and long ($>70$ day) rotation periods; from galactic kinematics, the transition is inferred to occur at ages of a few billion years on average, although it is mass dependent \citep{Newton2016, Newton2018}.

Even within a mode, gyrochronological principles may not apply; that is, a more slowly rotating M dwarf may be younger than a more rapidly rotating one. In \citet{Pass2022}, we studied M dwarfs in wide binaries with stars of known age in an attempt to probe the time dependence of spindown. We found that while there is gradual spindown within the rapidly rotating mode over a few billion years -- which would be promising for gyrochronology -- the dispersion in initial rotation rates is similarly large, likely stemming from differences in the disk lifetime and subsequent disk-locking early in the star's life \citep[e.g.,][]{Rebull2018}. A fully convective M dwarf with a 5-day rotation period could therefore be very young with slow initial rotation, or it could be a few gigayears old and gradually spinning down from a faster initial rate. In that work, we also found that the age of transition between modes can vary greatly from star to star, with some 0.2--0.3M$_\odot$ M dwarfs making the jump by 600Myr and others remaining rapidly rotating for gigayears. There is therefore no guarantee that a specific star in the slowly rotating mode is older than a specific star in the rapidly rotating one, although the slowly rotating population is older on average.

While the spindown of fully convective M dwarfs is complicated, understanding this process is a worthwhile pursuit, particularly in the context of exoplanets: fully convective M dwarfs host the only terrestrial exoplanets whose atmospheres are amenable to characterization with current and near-future instrumentation, such as JWST and the ELTs \citep[e.g.,][]{Snellen2013, Lovis2017, Morley2017}. Stellar rotation and activity are closely correlated \citep{Kiraga2007}, including X-ray emission \citep{Wright2011, Wright2018}, H\textalpha\ emission \citep{West2015, Newton2017}, UV emission \citep{France2018}, and flare rate \citep{Medina2020, Medina2022a}. An M dwarf that remains in the rapidly rotating mode for gigayears therefore subjects its attendant planets to a violent environment of high-energy photons and, likely, a corresponding high flux of charged particles for gigayears, decreasing the feasibility of atmosphere retention \citep[e.g.,][]{Lammer2007, Tilley2019}. A robust understanding of the spin and activity evolution of fully convective M dwarfs is therefore necessary to accurately model the evolution of planetary atmospheres.

Stellar mass is known to have a significant impact on the age at which a fully convective M dwarf transitions between rotation/activity modes \citep{West2008, Newton2016, Newton2017, Newton2018, Medina2022a}; to what extent do other variables play a role? While we attempted to investigate this question in \citet{Pass2022} by studying the rotation of M dwarfs in wide binary pairs with other M dwarfs, our conclusions were limited by selection biases, as it is much easier to measure short rotation periods than long ones.

In this work, we take a different approach to avoid incompleteness. As mentioned above, activity is an excellent proxy for rotation: using a volume-complete survey of single, low-mass M dwarfs, we found that 92$\pm$3\% of 0.1--0.3M$_\odot$ M dwarfs that are active in H\textalpha\ have rotation periods shorter than 20 days \citep{Pass2023}, with the remaining 8$\pm$3\% also having shorter rotation periods than the typical $>$100-day periods of slow rotators (see Figure 5 of \citealt{Pass2023}). By measuring H\textalpha\ emission for each component in wide M-M binary pairs, we can therefore probe the mass dependence of stellar spindown: does the more massive component always spin down first, or are there other factors at play? This observing strategy also controls for some variables, as binary stars likely formed together and therefore share their metallicity \citep{Desidera2004, Desidera2006} and birth environment.

In Section~\ref{sec:obs}, we discuss our observing strategy: Section~\ref{sec:targets} presents our target selection with Gaia, Section~\ref{sec:fast} our initial observing campaign with the mid-resolution FAST spectrograph, Section~\ref{sec:TRES} our follow-up campaign with the higher-resolution TRES spectrograph, and Section~\ref{sec:triple} the characterization of an intriguing new triple system. We discuss our results in Section~\ref{sec:discussion} and conclude in Section~\ref{sec:conclusion}.

\section{Observations}
\label{sec:obs}

\subsection{Target selection}
\label{sec:targets}
To select our targets, we conduct a search for common proper motion (cpm) pairs within 50pc from Gaia EDR3 \citep{GaiaCollaboration2016, GaiaCollaboration2021, Lindegren2021}, following a similar procedure to that outlined in \citet{Pass2022}. That is, we cross-match with the 2MASS catalog \citep{Cutri2003}, discard sources with absolute $K$-band magnitudes outside the range appropriate for M dwarfs ($5 < M_K < 10$ mag), and estimate masses for each remaining source using the \citet{Benedict2016} $K$-band mass-luminosity relation. This relation has an rms scatter of $\pm$0.014M$_\odot$. We identify cpm pairs using the proper motion ratio and proper motion position angle difference cuts of \citet[][i.e., a threshold of 0.15$^2$ in their Equation 1 and 15\textdegree\ in their Equation 2]{Montes2018}. We also require parallax agreement within 0.4mas; this is a slightly tighter constraint than the 2mas threshold used in \citet{Pass2022}, as we found in that work that such a lenient cut results in false positives in the form of single Hyads.

We further limit our list of cpm pairs given the science goals of this investigation. As we are interested specifically in fully convective M dwarfs, we remove sources with estimated masses outside the range 0.08--0.35M$_\odot$. We also remove one target that we identify as a white dwarf based on its $G_{\rm BP} - G_{\rm RP}$ color. From inspection of previous measurements from \citet{Newton2017}, we assume sources fainter than $m_R=15.5$ mag will be unsuitable for H\textalpha\ analysis with FAST; we therefore also limit our search to sources brighter than this threshold, with $R$-band magnitudes estimated using the empirical $G-K$ color relation from \citet{Winters2021}. We require that the components have separations between 4" and 2000", with the lower threshold representing our ability to resolve the targets with FAST under typical seeing and the upper boundary intended to mitigate false positives at extremely wide separations. Note that our results are insensitive to the exact choice of upper boundary, as we do not identify any pairs with separations between 1000--2000". We cut stars with Gaia EDR3 renormalised unit weight error (RUWE) values greater than 2; this quantity represents the excess noise in the Gaia astrometric solution, with values substantially larger than 1 suggesting that the star is likely an unresolved binary. We also only consider targets above declination -15\textdegree\ to ensure all sources are easily accessible from our telescope.

This search yields 66 pairs. We also manually add LHS 3808 / LHS 3809 to our target list, for a total of 67 pairs. These stars are missing from the gaiadr2.tmass\_best\_neighbour crossmatch table \citep{Marrese2019} that we use to obtain $K$-band magnitudes and therefore were not found by the above algorithm, but they nonetheless meet the criteria outlined above.

\subsection{FAST observations}
\label{sec:fast}
Our goal is to measure the H\textalpha\ feature for each component in our 67 wide-binary, mid-to-late M dwarfs pairs, and thus identify informative ``transition" systems in which one star is active and the other inactive. To this end, we observed each star with the FAST spectrograph \citep{Fabricant1998} at the 1.5m Tillinghast Reflector at the Fred Lawrence Whipple Observatory, with our observing campaign beginning in 2022 May and concluding in 2023 November. We reduced the spectra using the instrument's standard pipeline \citep{Tokarz1997}. Our campaign used the same settings as a previous search for H\textalpha\ emission in a different sample of M dwarfs, described in \citet{Newton2017}. Specifically, we attained roughly $R=3000$ resolving power over a wavelength range of  5550–7550\AA. We selected exposure times with the goal of attaining a per-pixel SNR of 40 in the continuum near the 6563\AA\ H\textalpha\ feature to ensure a clear detection of emission, if present.

We measure the equivalent width of the H\textalpha\ feature using the method described in \citet{Newton2017}, defining the feature as the wavelength range 6558.8–6566.8\AA\ and the continuum regions as 6500–6550\AA\ and 6575–6625\AA. We adopt the convention that a negative equivalent width indicates emission.

\citet{Newton2017} used -1\AA\ as their threshold to distinguish between active and inactive M dwarfs from FAST spectra, although a related study with the same instrument placed the boundary at -0.75\AA\ \citep{West2015}. \citet{Newton2017} noted that there were few stars in their sample with equivalent widths between \hbox{-0.5\AA} and -1.5\AA\, and so the exact value selected for this threshold is not strongly motivated. We therefore classify our stars as follows:

\begin{itemize}
    \item \textbf{Inactive}: If both stars in a pair have H\textalpha\ in absorption or negligible H\textalpha\ emission (H\textalpha\ $>$ -0.5\AA), the pair is considered inactive. That is, these are likely older stars that have already spun down to slow rotation and magnetic quiescence.
    \item \textbf{Active}: If both stars have obvious H\textalpha\ emission (H\textalpha\ $<$ -1\AA), the pair is considered active. That is, these are likely young, rapidly rotating stars with substantial magnetic activity.
    \item \textbf{Candidate Transition}: Pairs that do not fall into either of the above categories are candidate transition systems. In some cases, one star is clearly active and the other inactive; in others, the equivalent widths are near the activity threshold and higher-resolution follow up would be beneficial to confirm whether an emission feature is present in one or both stars.
\end{itemize}

\subsection{TRES observations}
\label{sec:TRES}
As we showed in \citet{Pass2022}, hierarchical triples like GJ 1006 and GJ 1230 can masquerade as transition systems, where one component remains active at an advanced age because activity and rotation are maintained by tidal interactions with a close binary companion; \hbox{G 68-34} \citep{Pass2023b} is another example of such a system, where a spin-orbit synchronized pair of fully convective M dwarfs has maintained rapid rotation and H\textalpha\ activity for over 5Gyr due to binary interactions. Our target selection uses a cut on Gaia RUWE to avoid many unresolved binaries, but there are some binary configurations for which we would not expect an astrometric perturbation (e.g., equal-mass binaries or very short-period binaries).

We therefore follow up the eleven candidate transition systems using the higher-resolution ($R=44000$) TRES spectrograph at the same 1.5m telescope to confirm the H\textalpha\ emission levels observed by FAST and vet the stars for unresolved binarity. We search for double lines in both stars, and obtain a second TRES observation for the active star in each pair after a few days to check for radial-velocity variability. To reduce the TRES observations and extract these radial velocities, we use our mid-to-late M-dwarf pipeline described in \citet{Pass2023a}.

Our TRES follow up yields one intriguing new triple system, which we discuss in Section~\ref{sec:triple}. We reclassify one system as active, one as inactive, and confirm that the remaining eight are transition systems with no evidence for unresolved binary companions. The final classification criteria are therefore:

\begin{itemize}
    \item \textbf{Inactive}: Both stars have H\textalpha\ $>$ -0.5\AA.
    \item \textbf{Active}: Both stars have H\textalpha\ $<$ -1\AA.
    \item \textbf{Transition}: One star has H\textalpha\ $>$ -0.5\AA\ and the other has H\textalpha\ $<$ -1\AA.
\end{itemize}

We tabulate the 8 transition systems in Table~\ref{tab:transition}, the 22 active systems in Table~\ref{tab:active}, and the 36 inactive systems in Table~\ref{tab:inactive}, for a total of 66 pairs. The 67th system is the newly discovered triple and is discussed separately, below. The tables include the 2MASS IDs of the components, the distance from Earth, the angular separation between components, the masses and mass difference between components, the H\textalpha\ equivalent widths measured by FAST, and (where applicable) the H\textalpha\ equivalent widths measured by TRES. While we include our nominal measurement uncertainty for the FAST equivalent widths, note that astrophysical variation in this feature over time is likely to introduce substantially larger uncertainties \citep[e.g.,][]{Medina2022}.

\subsection{LDS 942: an intriguing new triple}
\label{sec:triple}
The cpm pair 2MASS J12565215+2329501 and 2MASS J12565272+2329506 was known to be a binary long before Gaia, appearing in the Luyten Double Star catalog as LDS 942A and B \citep{Luyten1969}. Following our target selection described in Section~\ref{sec:targets}, we identify this pair as comprising a roughly equal-mass binary (0.334M$_\odot$ and 0.324M$_\odot$), located at a distance of 27pc from Earth and with the components separated by 8", implying a projected physical separation of 220au. With FAST, we measure H\textalpha\ equivalent widths of \hbox{-0.731$\pm$0.016\AA} for A and \hbox{-4.516$\pm$0.028\AA} for B. We therefore flag the pair as a candidate transition system, causing us to pursue follow up observations with TRES.

Given our motivation set forth in Section~\ref{sec:TRES}, we thought it possible that LDS 942B might be an unresolved binary, with its H\textalpha\ activity explained by binary interactions between two close components. However, we did not observe double lines in the TRES spectra of this star, nor statistically significant RV variation in five epochs of observations taken over a span of two months. The five observations have a sample standard deviation of 75ms$^{-1}$, providing a tight constraint on the possible existence of close companions. The star is thus presumably single (exempting its wide companion, LDS 942A).

We also collected an observation of LDS 942A to verify the H\textalpha\ equivalent width we observed in the FAST spectrum and were surprised to find that the A component was a double-lined spectroscopic binary. We continued to collect a total of ten TRES spectra of A in 2023 March/April in order to determine the orbit of the newly discovered LDS 942AC.

\begin{deluxetable}{lrrrr}[t]
\tabletypesize{\footnotesize}
\tablecolumns{7}
\tablewidth{0pt}
 \tablecaption{\texttt{TODCOR} results for LDS 942A and C \label{tab:rv}}
 \tablehead{
 \colhead{\textbf{BJD}} & 
 \colhead{\textbf{RV\bm{$_{\rm A}$}}} &
 \colhead{\textbf{RV\bm{$_{\rm C}$}}} &
 \colhead{\vspace{-0.1cm} \bm{$h$}} &  
 \colhead{\bm{$t_{\rm exp}$}} \\[-0.2cm]
 \colhead{[d]} &
 \colhead{[kms$^{-1}$]} &
 \colhead{[kms$^{-1}]$} &
 \colhead{} &
 \colhead{[s]}}
\startdata
2460007.8388 & -4.254  & -15.311 & 0.833 &3600 \\
2460016.8462 & -46.525 &  33.585 & 0.889 &3600 \\
2460030.8208 &   0.623 & -20.402 & 0.857 &3600 \\
2460036.7726 & -16.468 &  -0.885 & 0.939 &3600 \\
2460042.7963 & -35.849 &  21.387 & 0.911 &1800 \\
2460045.8009 &   1.725 & -21.731 & 0.891 &1800 \\
2460053.7711 &   4.129 & -24.326 & 0.866 &1800 \\
2460057.8406 &  -3.051 & -16.583 & 0.845 & 1800 \\
2460062.8897 & -20.496 &   4.121 & 0.896 & 1950 \\
2460064.7879 & -34.034 &  19.180 & 0.918 & 1950
\enddata
\tablecomments{$h$ is the cross-correlation coefficient \newline}
\end{deluxetable}
\vspace{-0.4cm}

\begin{figure}[t]
    \centering
    \includegraphics[width=\columnwidth]{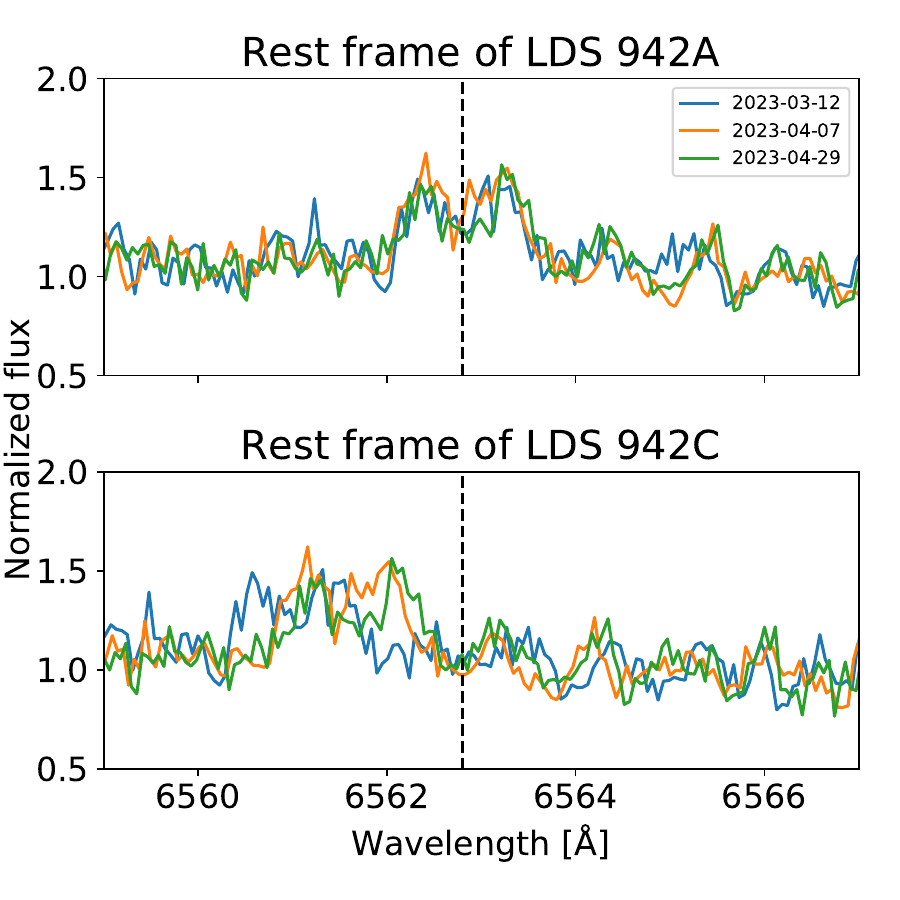}
    \caption{The 6563\AA\ H\textalpha\ feature (dashed line) in the blended spectrum of LDS 942AC, as observed by TRES. For clarity, we show only the epochs in which the lines of the two components are separated by at least 50kms$^{-1}$. In the upper panel, we have shifted the spectra to the rest frame of LDS 942A, and in the lower panel, to the rest frame of C. An H\textalpha\ emission feature is present in the spectrum of LDS 942A. The C component does not show any obvious emission.}
    \label{fig:ha}
\end{figure}

\begin{figure*}[t]
    \centering
    \includegraphics[width=0.95\textwidth]{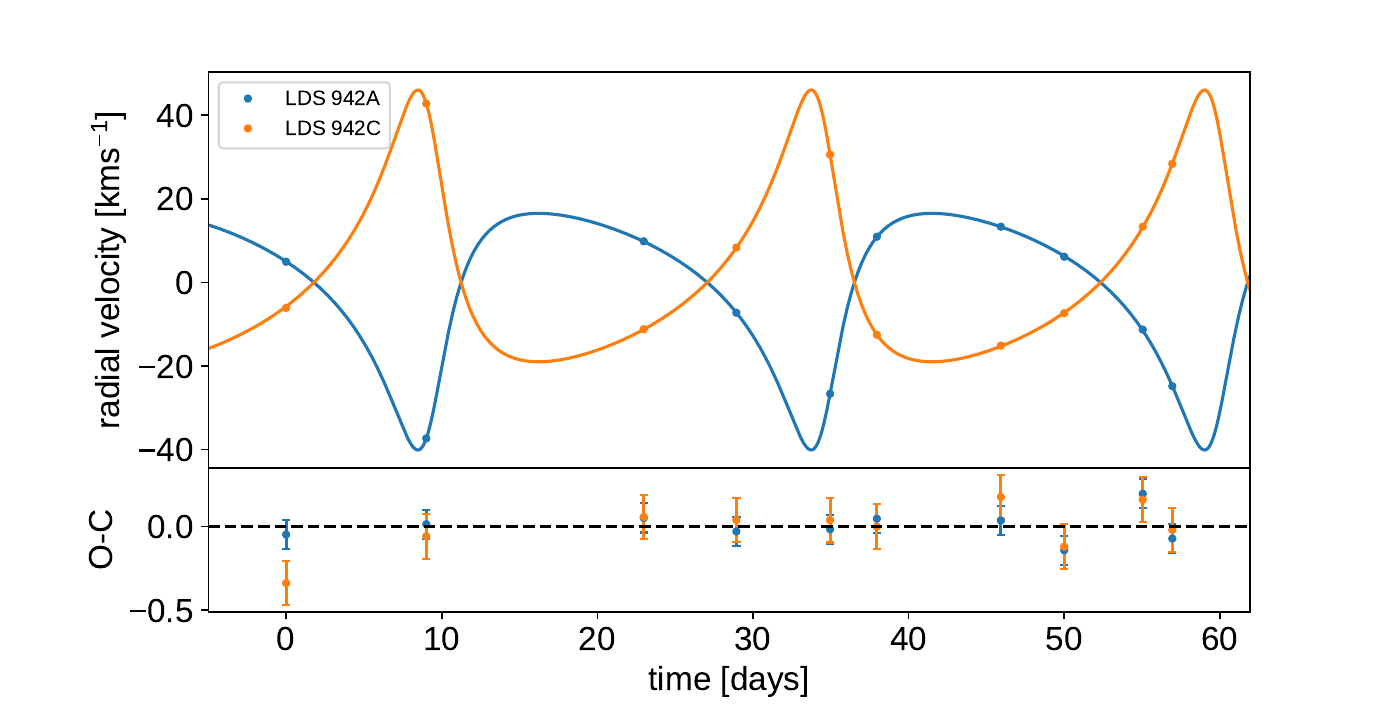}
    \caption{Our maximum \textit{a posteriori} orbital solution for the newly discovered close binary, LDS 942AC. The fit has a period of 25.27$\pm$0.02 days and a substantial eccentricity of 0.501$\pm$0.002.}
    \label{fig:LDS942AC}
\end{figure*}

We follow the method of \citet{Winters2020} to extract radial velocities of double-lined spectroscopic binaries from TRES spectra, which is based on the \texttt{TODCOR} technique \citep{Zucker1994}. This analysis uses TRES order 41, corresponding to wavelengths of 7065–7165\AA. As in \citet{Winters2020}, we use a spectrum of Barnard’s Star as the template for our cross correlation. We tabulate our extracted radial velocities in Table~\ref{tab:rv}. This analysis yields a C/A light ratio of 0.73 and no discernible rotational broadening for either star at the resolution of the spectrograph (i.e., $v$sin$i$ $<$ 3.4kms$^{-1}$). 

On the other hand, the spectra of LDS 942B are rotationally broadened, with $v$sin$i$ = 5.8kms$^{-1}$. LDS 942ABC was observed by the Transiting Exoplanet Survey Satellite (TESS; \citealt{Ricker2015}) in two sectors, with all three stars falling into the same TESS pixel; we observe a clear rotation period of 2.55 days in this blend and no other significant periodogram signals aside from harmonics of this period. Given our $v$sin$i$ measurements, LDS 942B is likely the source of this modulation: a mass of 0.324 M$_\odot$ would imply a radius of 0.321R$_\odot$ using the mass--radius relation of \citet{Boyajian2012}, yielding an equatorial velocity 6.4kms$^{-1}$ for a rotation period of 2.55 days. Considering an isotropic distribution of spin axes, the median value for sin$i$ is sin(60\degree)=0.87; this value would yield an observed rotational broadening of roughly 5.5kms$^{-1}$, close to the $v$sin$i$ we measure for LDS 942B. The rotation periods of the other two stars remain unknown.

Our TRES observations of LDS 942B yield a median H\textalpha\ equivalent width of -4.15\AA, measured following the method of \citet{Medina2020}. This value is comparable to the \hbox{-4.52\AA} equivalent width measured for this star from the FAST spectrum. For the blended LDS 942AC spectrum, the high resolving power of TRES allows us to measure the H\textalpha\ features individually for the two blended components. In Figure~\ref{fig:ha}, we show the three of our LDS 942AC spectra in which the two sets of lines are most separated. Inspection of these spectra reveals obvious H\textalpha\ emission in LDS 942A and no such signature in LDS 942C. Using these three spectra, we measure a median equivalent width of -0.83\AA\ for A in the blended spectrum; considering our \texttt{TODCOR} light ratio of 0.73, this measurement implies that the equivalent width would be -1.44\AA\ in a deblended spectrum of A. We note that if we take the median of our measurement for A in all ten spectra, we obtain a similar value of \hbox{-0.78\AA} in the blended spectrum; there therefore does not appear to be any additional H\textalpha\ emission introduced by component C, suggesting that C is H\textalpha\ inactive. We do not make a quantitative measurement of its H\textalpha\ equivalent width given the uncertainties introduced by the blend.

\begin{deluxetable}{lllll}[t]
\tabletypesize{\footnotesize}
\tablecolumns{5}
\tablewidth{0pt}
 \tablecaption{Orbital fit for LDS 942AC \label{tab:orbit}}
 \tablehead{
 \colhead{\textbf{Parameter}} & 
 \colhead{\textbf{Value}} &
 \colhead{\textbf{Error}} &
 \colhead{\textbf{Unit}}}
\startdata
$P$ & 25.274 &  0.016 & days \\
$T_{\rm peri}$ & 2460017.016 & 0.019 & BJD \\
$e$ & 0.5013 & 0.0020 & --- \\
$\omega$ & 214.01 & 0.27 & \degree \\
$\gamma$ & -9.22 & 0.50$^{\rm{*}}$ & kms$^{-1}$ \\
$K_A+K_C$ & 60.77 & 0.31 & kms$^{-1}$ \\
$M_A$sin$^3i$ & 0.2035 & 0.0028 & M$_\odot$ \\
$M_C$sin$^3i$ & 0.1772 & 0.0024 & M$_\odot$ \\
$q$ & 0.8707 & 0.0031 & --- \\
$\sigma_A$ & 0.129 & 0.031 &  kms$^{-1}$ \\
$\sigma_C$ & 0.160 & 0.025 &  kms$^{-1}$
\enddata
\tablecomments{*This error is dominated by the uncertainty in our determination of the absolute zero point of the TRES RV scale, resulting from the absolute RV error in a comparison spectrum of Barnard's Star \citep{Winters2020}. Relative RVs are known to much greater precision. \newline}
\end{deluxetable}
\vspace{-0.4cm}

\begin{figure*}[t]
    \centering
    \includegraphics[width=0.96\textwidth]{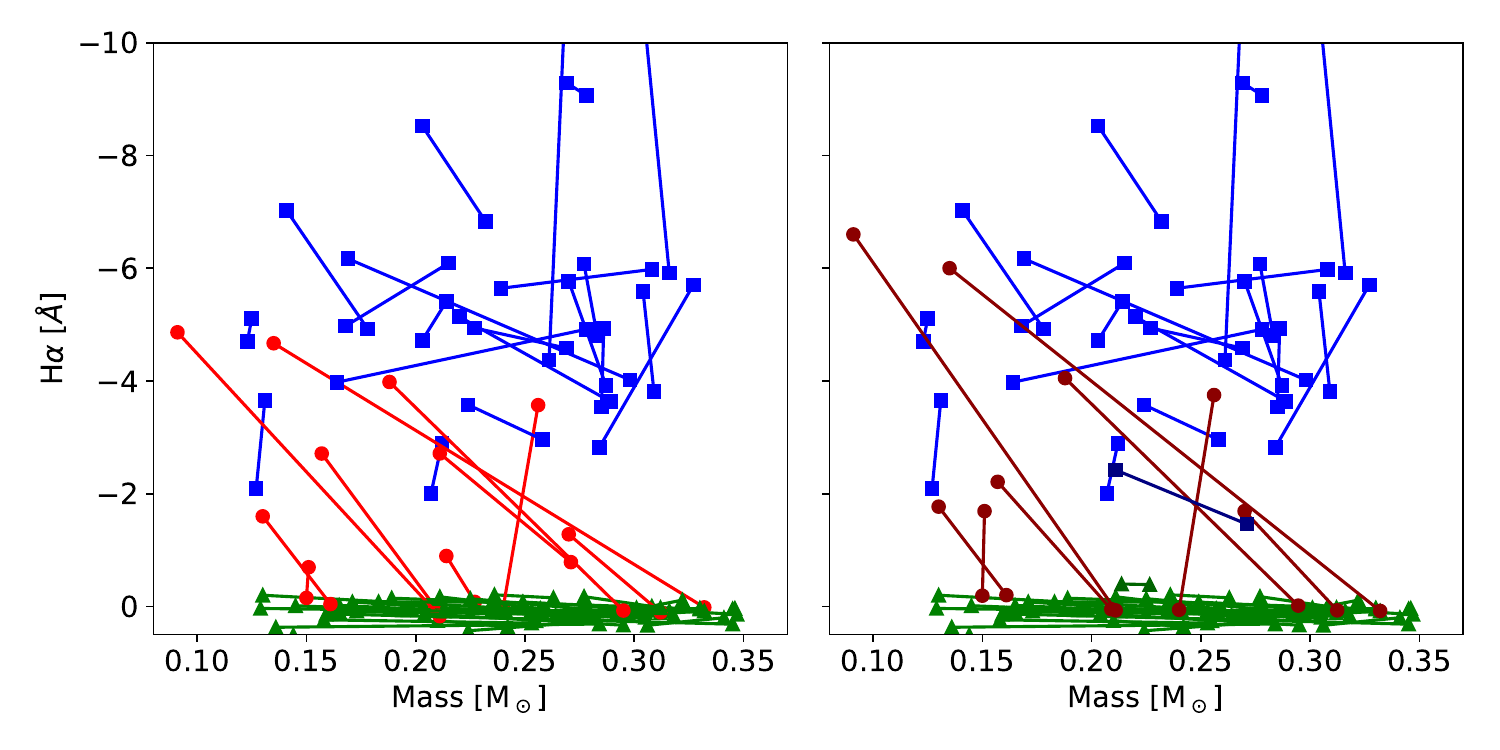}
    \caption{The H\textalpha\ equivalent widths for our 66 pairs, excluding the newly discovered triple. The two components are joined by a line. The left panel shows our original FAST observations. Active pairs are noted with blue squares, inactive pairs with green triangles, and candidate transition systems with red circles. In the right panel, we replace the measurements for the candidate transition systems with our follow-up observations from TRES, with red now denoting confirmed transition systems. We use a darker shade of each color in this plot to indicate the pairs with refined equivalent widths from TRES follow up.}
    \label{fig:pairs}
\end{figure*}

We use \texttt{exoplanet} \citep{ForemanMackey2021} to fit an orbital solution to our \texttt{TODCOR} radial velocities (Figure~\ref{fig:LDS942AC}). We allow the RV uncertainties to be free parameters of the fit, with each of the ten observations having the same error but with a different value for the two stellar components. We identify the maximum \textit{a posteriori} (MAP) solution with \texttt{exoplanet}, then use \texttt{PyMC3} \hbox{\citep{Salvatier2016}} to sample the posterior, starting from the MAP solution and using two chains each with a 1500 draw burn-in and 2000 draws. The orbital parameters inferred from this analysis are given in Table~\ref{tab:orbit}.

Our spectroscopic fit yields a mass ratio of $q=0.87$. To estimate the masses of the individual components, we assume the TRES light ratio approximates the light ratio in $R$ band. We use the deblending ratio from \citet{Riedel2014} to convert to a $K$-band light ratio, deblend the observed $K$-band magnitude into its two components, and apply the \citet{Benedict2016} relation to obtain component masses, yielding 0.23M$_\odot$ and 0.20M$_\odot$. These component masses are fully consistent with the spectroscopically determined mass ratio, implying an inclination of roughly 70\degree.

LDS 942B is therefore the brightest and most massive star in the system, with a mass of 0.32M$_\odot$. It is in the active phase of its life, with a short rotation period of 2.55 days and substantial H\textalpha\ emission (-4.15\AA). Its widely separated companion LDS 942A is fainter and less massive, with a mass of 0.23M$_\odot$, no observed rotational broadening, and modest H\textalpha\ emission (-1.44\AA). LDS 942C is a close companion to A and the faintest and least massive star in the system, with a mass of 0.20M$_\odot$. It does not show any rotational broadening and appears to be inactive in H\textalpha. The AC pair has a short and eccentric 25-day orbit.

Given our expectation that more massive M dwarfs spin down at younger ages, this system is unusual: the most massive component remains active and rapidly rotating, while the least massive component appears to have spun down (or at least, is magnetically quiescent). It is known that spin-orbit synchronization can lead to M dwarfs with anomalously long-lived rapid rotation and activity (e.g., \citealt{Pass2023b}); the LDS 942 system may hint that multiplicity can have even more insidious influences on the population of M dwarfs. That is, there may also be types of binary interactions that drive systems to longer rotation rates and magnetic inactivity. For example, \citet{Felce2023} discuss another spin-orbit equilibrium in which interactions from an outer companion drive an inner close binary into Cassini state 2, leading to very slow rotation. Future photometric monitoring of the LDS 942AC pair to determine the rotation periods of these components may therefore lead to new insights into the influence of stellar interactions on spindown and activity.

\begin{figure*}[t]
    \centering
    \includegraphics[width=\textwidth]{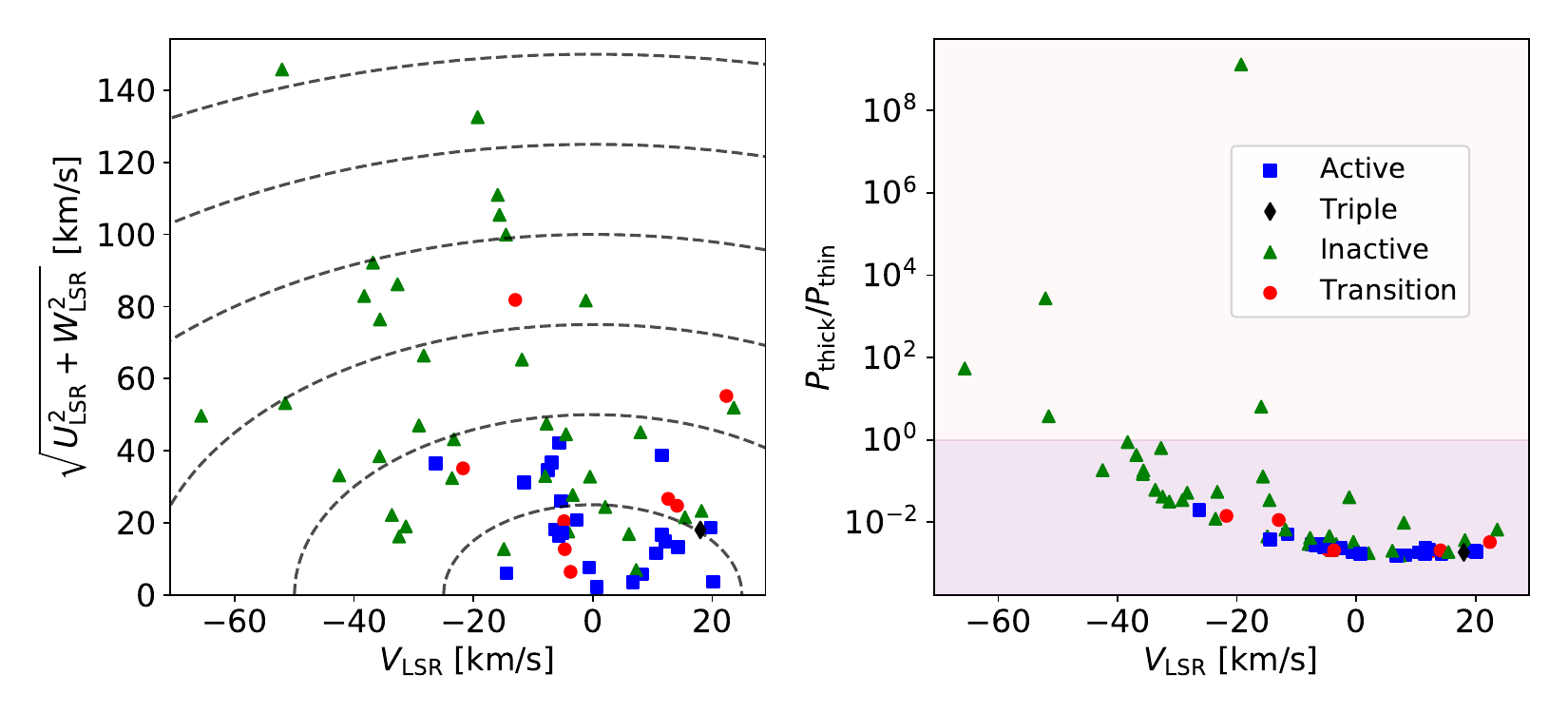}
    \caption{The galactic kinematics of our 67 pairs. In the left panel, we show a Toomre diagram, with the dashes indicating lines of constant total velocity. In the right panel, we show the likelihood that a star is a thin or thick disk member based on these kinematics; shaded regions indicate the regimes $P_{\rm thick}/P_{\rm thin} > 1$ and $P_{\rm thick}/P_{\rm thin} < 1$. The inactive population is dynamically hotter than the active population, in agreement with our expectation that the inactive stars are generally older. Our sample predominantly comprises thin disk stars.}
    \label{fig:galkin}
\end{figure*}

\section{Discussion}
\label{sec:discussion}
\subsection{Transition systems are less active}
Our H\textalpha\ equivalent widths for our 66 pairs (exempting the LDS 942 system) are shown in Figure~\ref{fig:pairs}. Inspecting the FAST observations in the left panel, we see that the active stars in the transition systems tend to have weaker H\textalpha\ emission than stars in active pairs: nearly half of the stars in active pairs have H\textalpha\ emission stronger than -5\AA\ as measured by FAST, while none of the stars in transition pairs do. This effect is less pronounced in the right panel, but that figure includes data from two different instruments; the larger H\textalpha\ features for the transition systems observed from TRES suggest that there may be some differences between the data sets, as different continuum regions are used to measure the feature given the dramatically different resolving powers of the two spectrographs.

In our investigation of a volume-complete sample of single, H\textalpha-active mid-to-late M dwarfs in \citet{Pass2023}, we observed a likely related phenomenon (Figure 6 of that work): stars with longer rotation periods within the rapidly rotating mode (periods of 2--10 days) tended to have lesser levels of H\textalpha\ emission than stars with the shortest rotation periods (periods less than 0.5 days), suggesting that H\textalpha\ emission tempers over time as M dwarfs spin down gradually within the rapidly rotating mode. This explanation suits our findings in this M--M binary sample: pairs in which one component has already spun down are likely to be older, and hence the remaining active component has had time to undergo substantial spindown, even though it has not made the jump to the slowly rotating sequence yet.

While our understanding of stellar spindown gives us good reason to assume that the transition pairs are younger on average than the inactive pairs and older on average than the active pairs, the galactic kinematics of the sample provide an independent test of this assertion. Using kinematics from Gaia and the method outlined in \citet[][which in turn follows from \citealt{Bensby2003}]{Medina2022a}, we calculate the UVW space motions of each star relative to the local standard of rest (Figure~\ref{fig:galkin}, left) and estimate disk membership based on these motions (Figure~\ref{fig:galkin}, right). From inspection of the Toomre diagram in this figure, we see that the population of active pairs is dynamically colder than the population of inactive pairs, with the transition pairs bridging the two populations. We can quantify this difference using an empirical relationship that links age to velocity dispersion in the direction of the Galactic north pole ($\sigma_{v_z}$); again, we follow \citet{Medina2022a}, which is based on the methods of \citet{Yu2018} and \citet{Lu2021}. Specifically, we use the age--velocity relation (AVR) from \citet{Lu2021}: $\rm{ln(age)} = \beta \mathrm{ln}\sigma_{\mathit{v_z}} + \alpha$, where $\beta=1.58\pm0.19$ and $\alpha= -2.80\pm0.53$. Treating the uncertainties in the AVR parameters as Gaussian, we estimate an average age of 1.4Gyr for the active subsample (68\% confidence interval of 0.7--2.7Gyr), 2.5Gyr for the transition subsample (1.2--4.9Gyr) and 10.6Gyr for the inactive subsample ($>$4.7Gyr). Despite large uncertainties in the kinematic age estimates, the inactive pairs are clearly older on average than the transition pairs, which are older on average than the active pairs.

As a related aside, inspection of the righthand plot in Figure~\ref{fig:galkin} shows that the sample overwhelmingly comprises thin disk stars; only five pairs in the sample have $P_{\rm thick}/P_{\rm thin} > 1$, indicating that they are more likely to be thick disk members than thin disk members, and only three of these have $P_{\rm thick}/P_{\rm thin} > 10$, which \citet{Medina2022a} require for a likely thick-disk classification. None of the stars have $P_{\rm halo}/P_{\rm thick} > 10$, which would indicate likely halo membership. The candidate thick-disk members are all inactive, which would be consistent with the older age of that population; \citet{Fantin2019} report that star formation rate was roughly uniform in the thin disk over the last 8Gyr, while star formation rate in the thick disk sharply peaked 10Gyr ago.

\subsection{A simple model of fully convective M-dwarf spindown}
\begin{figure}[t]
    \centering
    \includegraphics[width=\columnwidth]{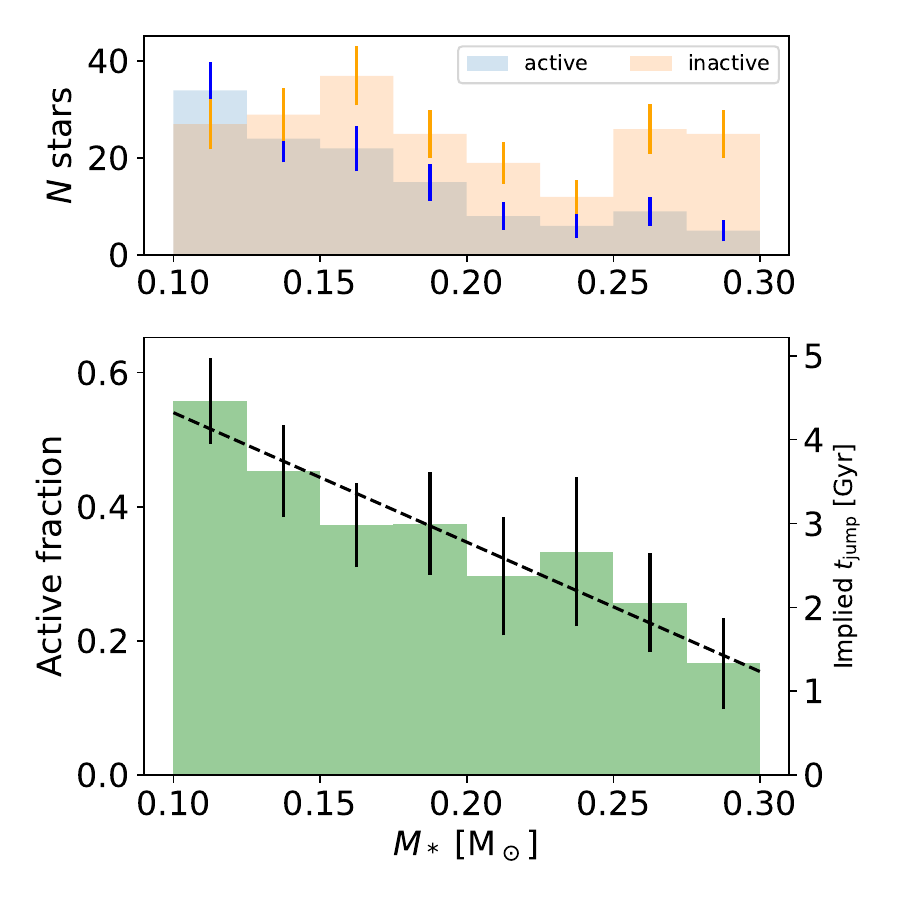}
    \caption{The upper panel shows a histogram of the single stars in the volume-complete sample of mid-to-late M dwarfs within 15pc \citep{Winters2021}, with the inactive sample taken from \citet{Pass2023a} and the active sample from \citet{Pass2023}. The lower panel shows the fraction of stars that are active. We find that this fraction is roughly linear with mass, following the trend $f=0.73-1.93M_*/M_\odot$. By assuming that star formation rate has been uniform in the solar neighborhood over the last 8Gyr (see text), the active fraction can be converted to a mass-dependent average age at which fully convective M dwarfs transition between activity and quiescence.}
    \label{fig:toy}
\end{figure}

While Figure~\ref{fig:pairs} is qualitatively interesting, we wish to make a more quantitative statement: is this distribution of pairs consistent with the epoch of spindown being determined solely by stellar mass? To answer this question, we can use the volume-complete sample of single, 0.1--0.3M$_\odot$ M dwarfs within 15pc \citep{Winters2021, Pass2023a, Pass2023} to construct a simple model of the mass dependence of spindown (Figure~\ref{fig:toy}). As this sample is volume complete, the active fraction as a function of mass is proportional to the average age of transition between modes, with the conversion requiring an assumption on star formation history. We follow \citet{Medina2022a} and assume that star formation has been constant over the last 8Gyr, which is motivated by the results of \citet{Fantin2019} for the galactic thin disk (note that our sample is predominantly thin disk stars, as discussed above).

Figure~\ref{fig:toy} shows that the M-dwarf active fraction -- and hence, the average age of transition -- scales roughly linearly with mass. For the active fraction, this is:
\begin{equation}
\label{eq:active}
f(M_*) = 0.73 - 1.93M_*/M_\odot,
\end{equation}
\noindent or for the age of transition:
\begin{equation}
\label{eq:epoch}
t_{\rm jump}(M_*) = 5.9 - 15.4M_*/M_\odot.
\end{equation}
\noindent A linear relationship implies that the probability of observing a pair in transition between the modes is proportional to the difference in mass between the components. By taking the derivative of Equation~\ref{eq:active}, we find that the probability of observing a system in transition is:
\begin{equation}
\label{eq:prob}
P(\Delta M_*)=0.193\frac{\Delta M_*}{0.1 M_\odot};
\end{equation}
\noindent i.e., there is a 19\% chance that a pair would be observed in transition if the component masses differ by 0.1M$_\odot$.

\begin{figure*}[t]
    \centering
    \includegraphics[width=\textwidth]{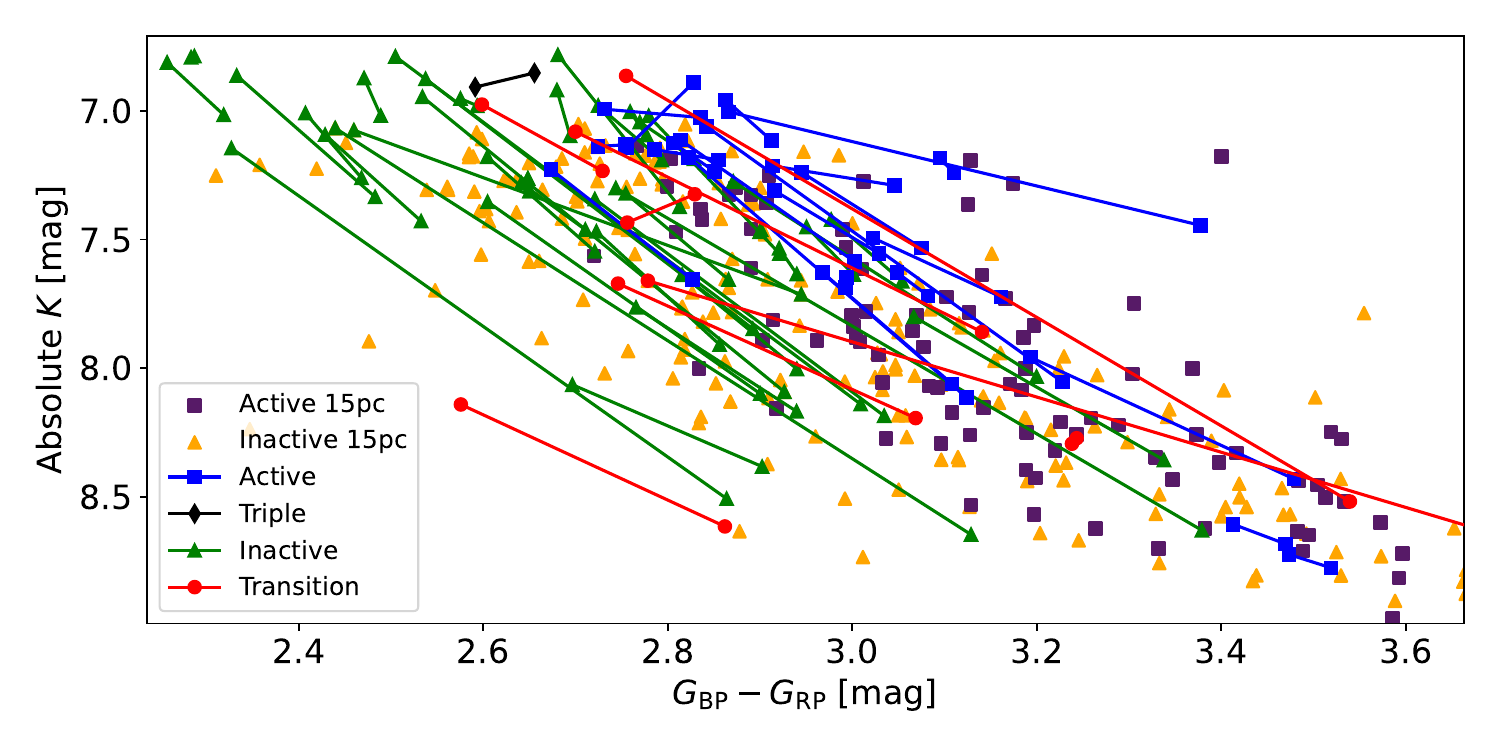}
    \caption{A color-magnitude diagram of our M-M binary pairs, overplotted on the volume-complete sample of single 0.1--0.3M$_\odot$ M dwarfs within 15pc \citep[from Figure 10 of][]{Pass2023}. The eight transition systems are shown in red. The axes provide the Gaia $G_{\rm BP}-G_{\rm RP}$ color and absolute $K$-band magnitude, which combines 2MASS apparent magnitude and Gaia parallax.}
    \label{fig:HRD}
\end{figure*}

\subsubsection{Evaluating the model at the population level}
Is our M-M binary sample consistent with this simple model? For each pair, we assign an age from a random uniform distribution spanning 0--8Gyr. For each component, we calculate $t_{\rm jump}$ given its mass; if the star is older than this value, we consider it inactive, and if it is younger, we consider it active. We repeat this simulation 1000 times, recording the number of active, inactive, and transition pairs that we obtain. This exercise yields 14.3$\pm$3.3 active pairs, 45.0$\pm$3.8 inactive pairs, and 6.7$\pm$2.3 transition pairs. While the number of transition pairs is in agreement, the simple model predicts significantly fewer active pairs and more inactive pairs than we observed (which were 22 and 36, respectively).

Note that while we performed these calculations in terms of age for ease of explanation, we could alternatively have done so entirely in terms of active fraction; the assumption we make about the star formation rate cancels out, as it is applied to both our equation for $t_{\rm jump}$ and our assigned ages. The discrepancy therefore cannot be explained by an inaccurate assumption of that history (although we are assuming that the star formation history for single stars within 15pc is the same as for binaries within 50pc).

A possible explanation for our overabundance of active systems is selection bias: our target selection is magnitude limited (unlike the volume-complete sample), and young stars are overluminous. To investigate this possibility, we rerun our Gaia target selection algorithm but remove the requirement that stars must be brighter than $m_R=15.5$ mag. This change would nearly double our sample size, meaning that a preferential selection of young, overluminous stars could have an impact on our sample. However, we would not expect the difference to be dramatic: if one assumes that these M dwarfs would be overluminous for 300Myr (see discussion in \citealt{Pass2022}) and star formation has been uniform for 8Gyr, 4\% would be overluminous; even if their overluminosity caused all of them to enter our sample, this bias would only result in three additional active systems. We also vet our sample for members of young moving groups using the \texttt{BANYAN} $\Sigma$ tool \citep{Gagne2018} and radial velocities, proper motions, and parallaxes from Gaia. All but two of the pairs are most likely field stars and therefore unlikely to be overluminous. One pair, 2MASS J03513447+0722250 and 2MASS J03513420+0722229, is likely a member of the Hyades with age 600--800Myr \citep{Brandt2015}; we also do not expect overluminosity at this age. The other pair has a high membership probability for the Carina-Near Moving Group, for which overluminosity would be expected, but as we discuss in Section~\ref{sec:stahl}, past work has argued that the Carina-Near classification is incorrect for this system and the pair actually belongs to the field.

Figure~\ref{fig:HRD} shows a color-magnitude diagram of our sample, overplotted on the volume-complete 15pc sample of single, 0.1-0.3M$_\odot$ M dwarfs from \citet{Pass2023a, Pass2023}. This figure is consistent with our discussion thus far: there is a handful of active systems whose positions could be consistent with overluminosity (located in the upper right of the figure), but most do not appear to be overluminous. The active pairs do tend to be redder than the inactive pairs at constant luminosity, but this is also observed in the volume-complete sample and may be the result of starspot coverage \citep[e.g.,][]{Covey2016}. To confirm that differences in starspot coverage could feasibly generate the offsets we observe, we examine the stellar evolutionary models of \citet{Somers2020}, which include the structural effects of starspots. We use $V-I_c$ as a proxy for $G_{\rm BP} - G_{\rm RP}$, as Gaia colors have not been calculated for low-mass M dwarfs in the \citet{Somers2020} grid for $f_{\rm spot} \neq 0$. For a 0.3M$_\odot$ M dwarf, these models predict a 0.54 mag difference between stars with $f_{\rm spot}=0$ and $f_{\rm spot}=0.85$, the minimum and maximum values modelled. A more modest difference in starspot filling fraction could therefore explain the observed data; for example, the $f_{\rm spot}=0.34$ and $f_{\rm spot}=0.68$ models are separated by an offset of roughly 0.2 mag.

Another possible explanation for our overabundance of active stars is unresolved binarity. While we have carefully vetted our transition systems for unresolved companions, the active pairs have not been followed up with a higher resolution spectrograph. \citet{Winters2019} showed that roughly 20\% of M-dwarf systems consist of a close binary with separation less than 50au, although this fraction does decrease for low-mass M dwarfs. Some of these close binaries would also produce astrometric perturbations and hence be rejected by our cut on Gaia RUWE; nonetheless, some would be missed, as with LDS 942AC. However, this effect could not turn an inactive system into an active system unless both components were unresolved binaries, which would be unlikely; the fact that we did not identify an abundance of unresolved binaries in our TRES follow-up of transition pairs therefore disfavors this hypothesis. On the other hand, binarity also acts to change the component masses; unresolved binarity could merely be leading to inflated mass estimates and hence underestimates of the active lifetime. That is, if we were to artificially decrease the masses of some of our stars, our simulation would predict a greater number of active pairs. However, the color-magnitude diagram in Figure~\ref{fig:HRD} also disfavors an abundance of unresolved binaries. While there is substantial thickness to the color-magnitude diagram at these low masses, the M-M binary pairs have similar slopes, which is unsurprising if the thickness of the color-magnitude diagram is caused by characteristics shared by both components of the binary, such as age or metallicity. Unresolved triples are likely to have unusual slopes, as potentially both the magnitude and the color of one of the points is incorrect due to the blend. The slopes of the active pairs generally appear similar to those of the inactive pairs (with a handful of exceptions: most notably, LP 196-29 / LP 196-30), again suggesting that an abundance of unresolved binaries is unlikely. And of course, unusual slopes do not necessitate binarity; if starspot coverage indeed has a significant impact on the CMD position of a given star, some differences between members of a binary pair might be expected even though metallicity and age are constant.

A remaining possibility is that Equation~\ref{eq:active} is different for binaries than it is for single stars. One reason this might occur is if the population of binaries is evolving with time. Such an outcome is expected given Heggie's Law: interactions with other stars will cause the orbit of a hard (close) binary to harden, and a soft (wide) binary to soften \citep{Heggie1975}. Hardening could cause binaries to evolve to separations smaller than our 4" lower limit, excluding them from our sample selection. Softening orbits could lead to the binary becoming unbound, thereby becoming single stars and also evading our target selection. Considering the density of the field environment, we expect our wide binaries to be soft. In this framework, Equation~\ref{eq:active} overpredicts the number of inactive stars we would observe because a subset of the predicted stars are no longer wide binaries in their old age. Alternatively, denser star formation environments in the past could disrupt binaries and cause older stars to have a lower primordial binary fraction \citep{Parker2014, Longmore2014}, although \citet{Moeckel2011} argue that the formation of soft binaries in clusters is actually independent of cluster size.

\begin{figure}[t]
    \centering
    \includegraphics[width=\columnwidth]{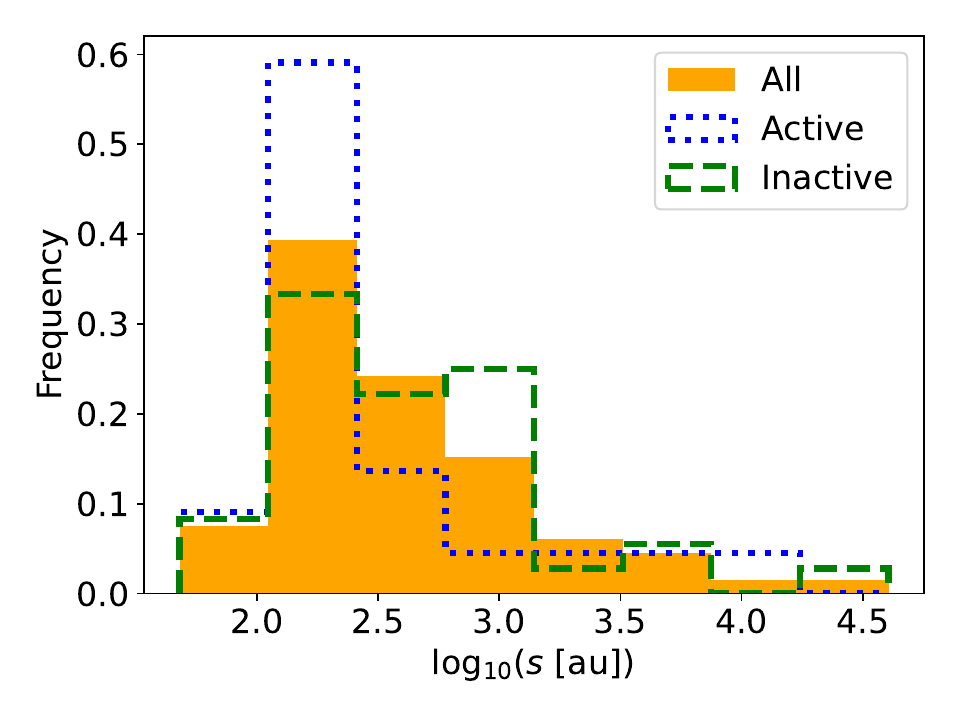}
    \caption{The projected physical separations between binary components. All 66 active, inactive, and transition pairs are included in the orange histogram; the blue dotted line indicates only the active pairs while the green dashed lines shows only the inactive pairs. A KS test yields tentative evidence (77\% confidence) that the active and inactive pairs are drawn from different distributions, with the active pairs having preferentially smaller separations.}
    \label{fig:sep}
\end{figure}

Do we see any indication of dynamical processing? In Figure~\ref{fig:sep}, we show a histogram of the projected separations for the active and inactive pairs. There is tentative evidence of a statistically significant difference between these populations, with a KS test yielding a 23\% chance that the two samples are drawn from the same distribution. The inactive pairs  tend towards larger separations than the active pairs. This result could be a signature of dynamical processing: the models of \citet{Jiang2010} that simulate the evolution of wide field binaries due to gravitational perturbations from passing stars find that these interactions shift the separation distribution to wider separations for stars that remain bound, and also result in many pairs becoming unbound. Dynamical processing could therefore explain both the shift in Figure~\ref{fig:sep} as well as the underabundance of inactive pairs we observe relative to our simple model expectations.

To approximate this effect in our simple model, we make the assumption that the probability of a binary experiencing a disruptive interaction is uniform in time. We therefore modify our age prior: instead of a flat distribution, there is a linear decrease in the likelihood of observing a binary pair at a given age. This treatment requires a normalization that represents the fraction of wide binary pairs that remain undispersed for 8Gyr. We test a variety of values. We find that if 20\% of wide binaries remain in existence for 8Gyr, then we would expect 21.2$\pm$3.8 active pairs, 8.8$\pm$2.6 transition pairs, and 36.0$\pm$4.0 inactive pairs, in agreement with our observations. As a statistical ensemble, our observations are therefore plausibly consistent with a picture in which stellar mass alone determines the age at which a fully convective M dwarf spins down, at least when metallicity and stellar birth environment are controlled.

To investigate the alternative, we also consider adding a random dispersion to our spindown ages. We model this effect by adding an offset drawn from a random normal distribution with $\sigma$=1Gyr to Equation~\ref{eq:epoch}, with each component receiving a different offset. We still include the dynamical processing described above. This treatment predicts an increase to 14.5$\pm$3.4 transition systems, which is disfavored by our observations. A dispersion of 0.5Gyr would yield 10.9$\pm$2.9 transition systems, which remains plausible at 1$\sigma$. If we instead model the dispersion as a fractional effect, a 25\% dispersion in the epoch of spindown also produces agreement within 1$\sigma$. In summary, our population-level observations are consistent with either no or modest dispersion.

\subsubsection{Evaluating the model for specific systems}
The individual transition systems can also grant us additional insights. Two of our observations are nominally impossible given our simple model: 2MASS J11231269+8009027 / 2MASS J11231650+8009045 and 2MASS J10204884-0633195 / 2MASS J10205111-0634400, where the less massive component has spun down first. In both cases, the masses of the two components are similar. Considering the $\pm0.014$M$_\odot$ uncertainty in the \citet{Benedict2016} mass--luminosity relation, it is plausible that we are mistaken in our identification of the more massive component, particularly for 2MA1123+80AB in which the masses are nearly identical. However, the mass difference for 2MA1020-06AB is 0.016M$_\odot$; while this is only marginally larger than the the nominal error in the mass--luminosity relation, \citet{Benedict2016} note that this scatter likely stems from a combination of age, metallicity, and magnetic effects. Our binary pairs presumably share a common age and metallicity, and hence a lesser degree of scatter should be expected than in the relation at large.

\begin{figure}[t]
    \centering
    \includegraphics[width=\columnwidth]{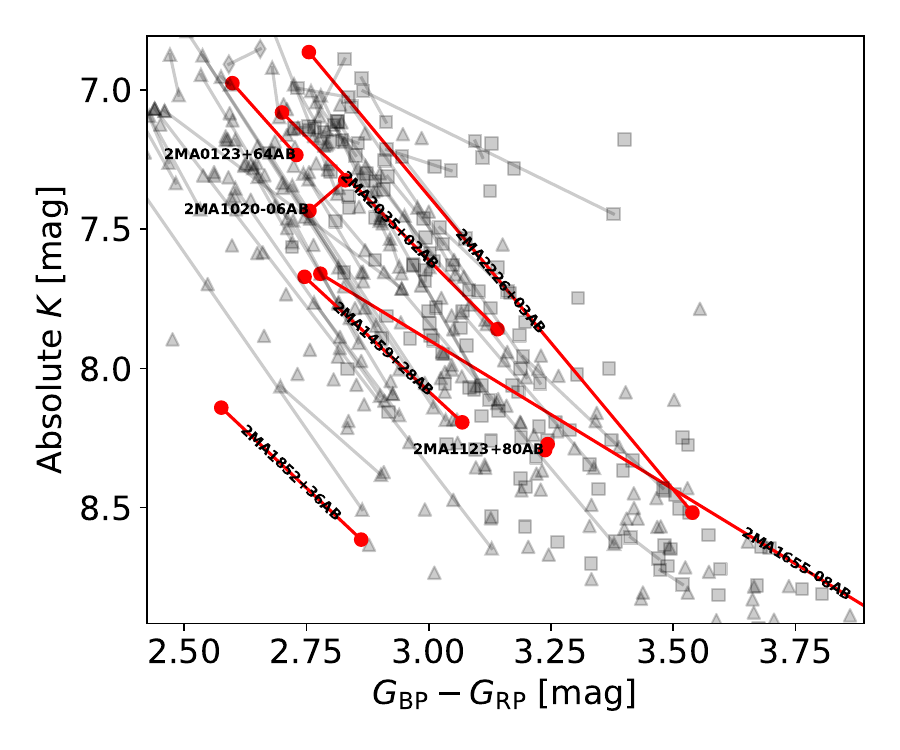}
    \caption{This plot shows the same data as Figure~\ref{fig:HRD}, but with the transition systems annotated.}
    \label{fig:HRD2}
\end{figure}

\begin{figure*}[t]
    \centering
    \hspace{-1cm}
    \includegraphics[width=1.03\textwidth]{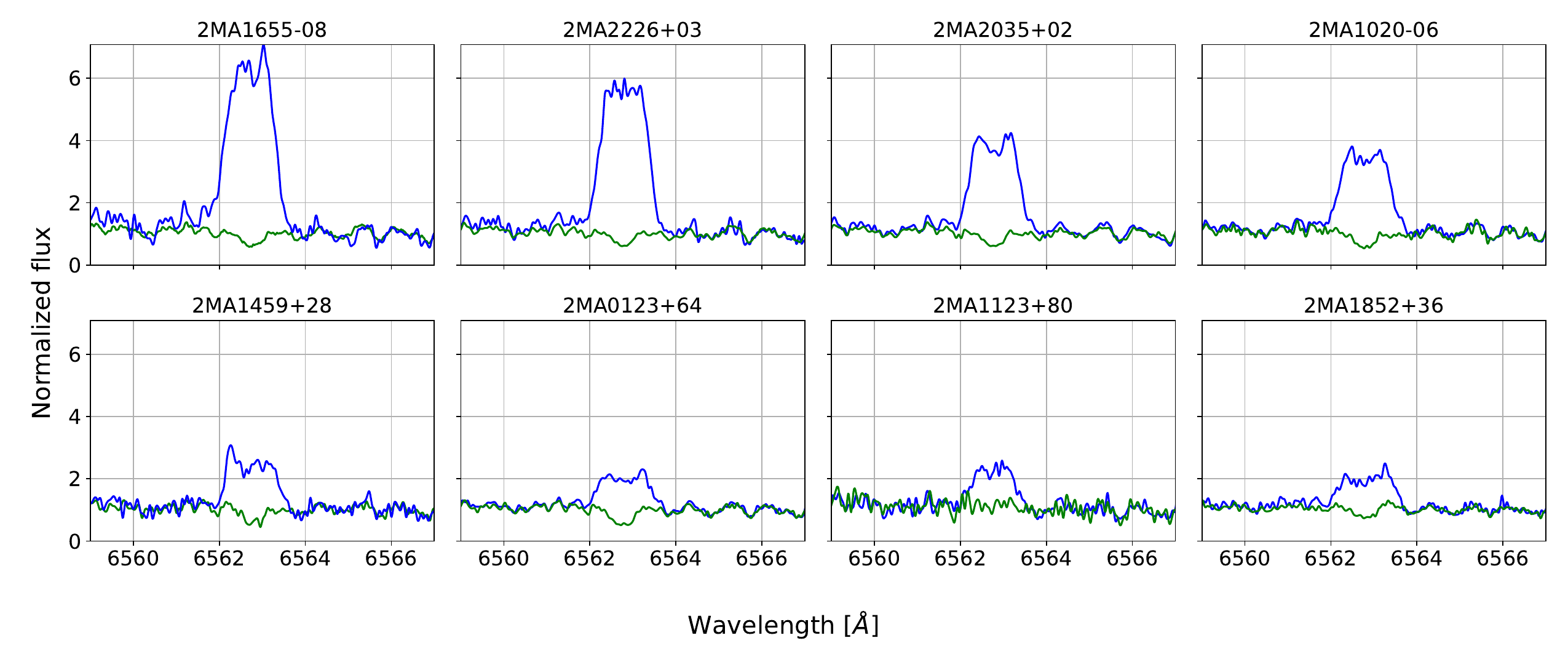}
    \caption{The 6563\AA\ H\textalpha\ feature for our transition pairs, as observed by TRES. The active component is shown in blue and the inactive component in green. Equivalent widths are tabulated in Table~\ref{tab:transition}.}
    \label{fig:bin_ha}
\end{figure*}

Magnetic effects do remain a consideration, although starspot-induced photometric modulation is unlikely to cause the discrepancy: fully convective M dwarfs that are modestly active or inactive have typically photometric peak-to-peak amplitudes of 0.01 mag in the optical \citep{Pass2023}, with even smaller amplitudes expected at longer wavelengths \citep[e.g.,][]{Miyakawa2021}. However, differing filling factors of longitudinally homogeneous spots (which do not cause photometric time variation) could play a role \citep[e.g.,][]{Irwin2011, Jackson2013}. We note that for both of these pairs, the brighter/active component is redder (Figure~\ref{fig:HRD2}), a behavior we see generally for both the active stars in this study and for the active single stars in the volume-complete sample from \citet{Pass2023}. On the other hand, work such as \citet{Morrell2019} suggests that there is no statistically significant difference between active and inactive M dwarfs with respect to either starspot coverage or radius inflation, which would disfavor this explanation.

Even if the inactive component in these pairs is more massive, the mass differences are small. While it would therefore be possible to observe these systems in transition given our simple model framework, is it likely? To investigate this question, we consider all pairs with mass differences of $\Delta M_* < 0.02$M$_\odot$. We assume that we cannot know a mass difference to better than this threshold due to the systematics described above, and hence the probability of observing any of these 18 roughly equal-mass pairs in transition is equally likely. We substitute 0.02M$_\odot$ for $\Delta M_*$ into Equation~\ref{eq:prob}, yielding 3.86\% as a conservative estimate of the likelihood of each of these systems being observed in transition; note that this corresponds to a difference in spindown epoch of 0.31Gyr. Using binominal statistics, we find an 85\% chance that we would have observed fewer transition pairs among the equal-mass binaries than the two that we did. While stellar mass is therefore the dominant effect in determining when a given star spins down (when metallicity and birth environment are controlled), our results hint that second-order effects may also be present; this could be explained by, for example, a 0.5Gyr dispersion in spindown epoch, which we found was plausibly consistent with our population-level results in the previous section. That said, there is still a 15\% chance that the data are consistent with the simple model; however, this agreement requires the assumption that we cannot know a mass to better than 0.02$M_\odot$, and a less conservative assumption would worsen the disagreement.

\subsection{Commentary on the transition systems}
The eight transition systems are exciting targets for future follow up work, representing pre- and post-spindown laboratories that are controlled for confounding variables such as differences in metallicity, birth environment, and age. Two of the systems are equal-mass binaries and hence even more exciting, as they are also controlled for stellar mass. In this section, we summarize the current state of knowledge on these systems; Figure~\ref{fig:bin_ha} shows our TRES observations of H\textalpha.

\subsubsection{2MASS J11231269+8009027A and 2MASS J11231650+8009045B}

2MA1123+80AB is an equal-mass pair with masses of 0.15M$_\odot$, separated by 10" and located at 26pc, which yields a projected physical separation of 260au. We do not observe rotational broadening in our TRES spectra of either star, implying $v$sin$i$ $<$ 3.4kms$^{-1}$. We observe H\textalpha\ emission of A in our FAST spectrum and two TRES spectra, taken in 2023 January, February, and April, respectively; it is therefore unlikely that we simply caught a quiescent star during a flare on three separate occasions. The pair appears at nearly identical magnitudes and colors in Figure~\ref{fig:HRD2}, with the A component being marginally brighter and redder.

From our TRES observations, we measure radial velocities of \hbox{-17.0$\pm$0.5kms$^{-1}$} for A and \hbox{-17.9$\pm$0.5kms$^{-1}$} for B. The errors in these measurements are dominated by the uncertainty in absolute RV of a Barnard's Star template that we use to calibrate our absolute RV scale; the differences between A and B are therefore statistically significant and likely reflect the mutual acceleration of the binary. We do not observe statistically significant variation between our two observations of A that are separated by two months and have relative RV uncertainties of 60ms$^{-1}$.

While the pair has been observed in many TESS sectors, we are unable to identify one or more rotation periods from the blended light curve.

\subsubsection{2MASS J10204884-0633195A and 2MASS J10205111-0634400B}
\label{sec:stahl}
2MA1020-06AB is a nearly equal-mass pair with masses of 0.26M$_\odot$ and 0.24M$_\odot$. The components are separated by 87" and located at 31pc, yielding a projected physical separation of 2700au. The system was previously identified as a cpm pair in \citet{Boyd2011}, who refer to the pair as SCR J1020-0633A and SCR J1020-0634B.

Neither star shows rotational broadening in our TRES spectra, implying $v$sin$i$ $<$ 3.4kms$^{-1}$. The stars are resolved separately by TESS; inspecting the TESS light curve of 2MA1020-06A, we find a rotation period of 3.8 days. We are unable to identify a period in the light curve of B. Our photometric and spectroscopic observations are consistent: using the \citet{Boyajian2012} mass--radius relation, we predict an equatorial velocity of 3.6kms$^{-1}$ for A given a 3.8-day rotation period, which would result in undetectable rotational broadening for most values of sin$i$. We measure a radial velocity of 16.4$\pm$0.5kms$^{-1}$ for A and 16.3$\pm$0.5kms$^{-1}$ for B.

Intriguingly, the \texttt{BANYAN} $\Sigma$ tool \citep{Gagne2018} finds that these stars' galactic space motions are consistent with the Carina-Near Moving Group, with a membership probability of 99.7\%. This result was previously reported in \citet{Stahl2022}, who also measured H\textalpha\ equivalent widths for these stars and obtained measurements consistent with our findings. Carina-Near has an age of 200Myr \citep{Zuckerman2006}; such a young age for 2MA1020-06B would be incredibly surprising given its H\textalpha\ absorption. For this reason, \citet{Stahl2022} argue that this star is not a true member of Carina-Near despite its high membership probability. Instead, they suggest that its misclassification is the result of the kinematics of Carina-Near being poorly defined in \texttt{BANYAN} $\Sigma$, as only 13 members of this association were previously known. \citet{Stahl2022} do not acknowledge the binarity of 2MA1020-06AB and so they still consider 2MA1020-06A to be a Carina-Near member, as it does have H\textalpha\ emission, but the same misclassification argument would apply to both stars.

\subsubsection{2MASS J18524373+3659257A and 2MASS J18524397+3659176B}

2MA1852+36AB consists of stars with masses of 0.16M$_\odot$ and 0.13M$_\odot$, separated by 9" at a distance of 23pc. These values imply a physical projected separation of 210au. A is inactive in H\textalpha\ while the B component is active.

We observe a weak 1.7-day rotation period in the blended TESS light curve, presumably originating from the active B component. Neither star exhibits rotational broadening in our TRES spectra, but a non-detection could be consistent with a 1.7-day rotation period for B if $i<40$\degree. We measure a radial velocity of -19.4$\pm$0.5kms$^{-1}$ for A and -19.8$\pm$0.5kms$^{-1}$ for B.

\subsubsection{2MASS J01232866+6411443A and 2MASS J01233090+6411440B}
This system is a known cpm pair, referred to as UC 13 in the USNO CCD Astrographic Catalog \citep[UCAC;][]{Caballero2010, Hartkopf2013}. It comprises 0.31M$_\odot$ and 0.27M$_\odot$ stars separated by 15". At a distance of 32pc, this corresponds to a physical projected separation of 480au. A is inactive in H\textalpha\ while the B component is active.

Neither star shows rotational broadening in our TRES spectra, nor do we observe a rotation period in the blended TESS light curve. We measure radial velocities of -7.3$\pm$0.5kms$^{-1}$  for A and -7.7$\pm$0.5kms$^{-1}$ for B.

\subsubsection{2MASS J14593063+2833387A and 2MASS J14593085+2833463B}
This pair appears in the Luyten Double Star catalog as LDS 6302AB \citep{Luyten1995}. It consists of 0.21M$_\odot$ and 0.16M$_\odot$ stars, separated by 8" and at a distance of 33pc, yielding a projected physical separation of 260au. A is inactive in H\textalpha\ while the B component is active.

Like the previous system, neither star shows rotational broadening nor do we observe a rotation period in the blended TESS light curve. We measure radial velocities of 11.8$\pm$0.5kms$^{-1}$  for A and 11.6$\pm$0.5kms$^{-1}$ for B.

\subsubsection{2MASS J20350677+0218289A and 2MASS J20350608+0218166B}
This pair appears as a cpm pair in the UCAC with designation UC 4224AB \citep{Hartkopf2013}. The components have masses of 0.29M$_\odot$ and 0.19M$_\odot$. They are separated by 16" and located at 20pc, yielding a projected physical separation of 320au. A is inactive in H\textalpha\ while the B component is active.

We again do not observe any rotational broadening in our TRES spectra nor a rotation period in the blended TESS light curve. We measure radial velocities of 7.0$\pm$0.5kms$^{-1}$ for A and 7.4$\pm$0.5kms$^{-1}$ for B.

\subsubsection{2MASS J16552527-0819207A and 2MASS J16553529-0823401B}

Our penultimate transition pair is known by many names, including Gl 643 and Gl 644C. This pair is part of a quintuple star system, with both components widely separated from each other and from the close triple, Gl 644, which is itself composed of three M dwarfs \citep{Mazeh2001}. We did not observe the triple in this study.

The two stars we did observe have masses of 0.21M$_\odot$ and 0.09M$_\odot$ and are separated by 299". As they are located at only 6pc, this corresponds to a projected physical separation of 1800au. Gl 643 is inactive in H\textalpha\ while Gl 644C is active.

The pair has not been observed by TESS, but \citet{DiezAlonso2019} report a rotation period of 6.5 days for the more massive component based on ASAS photometry \citep{Pojmanski1997}. Such a short rotation period would be highly unusual for a fully convective M dwarf with H\textalpha\ in absorption, with rotation periods of 95$\pm$22 days being typical for inactive M dwarfs at this mass \citep{Newton2017}, but perhaps could reflect the special phase of the star's life in which we are observing these transition systems. We do not observe a signal with this rotation period in 4559 observations from the MEarth Project \citep{Nutzman2008, Irwin2015} taken between 2021 February--July, although we are also unable to identify a different period. Further observation of this target would be beneficial to establish whether it is truly a rapid rotator.

While Gl 643 does not exhibit rotational broadening in our TRES spectra, we measure $v$sin$i$ of 6.1kms$^{-1}$ for Gl 644C. This broadening implies that the rotation period of this star is shorter than 1.2 days. A comparable value of $v$sin$i$=5.4$\pm$1.5kms$^{-1}$ was previously measured for this star with the higher-resolution ($R=94600$) CARMENES-VIS  spectrograph in \citet{Reiners2018}. We observe a tentative 1.095-day rotation period in 3754 MEarth observations taken between 2014--2018, consistent with the observed rotational broadening, although this signal did not pass the significance threshold to be considered a robust detection in \citet{Newton2018}.

We measure a radial velocity of 16.0$\pm$0.5kms$^{-1}$ for Gl 643 and 14.5$\pm$0.5kms$^{-1}$ for Gl 644C.

\subsubsection{2MASS J22261576+0300182A and 2MASS J22261549+0300075B}
The transition pair with the largest mass difference between components is 2MA2226+03AB, better known as LHS 3808 and LHS 3809, and designated as LDS 4967AB in the Luyten Double Star catalog \citep{Luyten1995}. The primary has a mass of 0.33M$_\odot$ and the secondary has a mass of 0.14M$_\odot$. The components are separated by 12", which at a distance of 23pc corresponds to a projected physical separation of 280au. This is the transition pair with the largest total velocity in Figure~\ref{fig:galkin}, suggesting it may have an age comparable to the population of inactive pairs.

We discussed this pair in a previous investigation, \citet{Pass2022}, as both components have rotation periods measured from MEarth photometry \citep{Newton2016}: 94 days for the inactive LHS 3808 and 1.6 days for the active LHS 3809. With our new TRES spectra, we measure rotational broadening of $v$sin$i$=5.0kms$^{-1}$ for LHS 3809 and no measurable broadening for LHS 3808 at the resolution of the spectrograph. A 1.6-day rotation period for LHS 3809 yields an equatorial velocity of 5.8kms$^{-1}$, consistent with our $v$sin$i$ measurement for a modest inclination. We measure RVs of \hbox{-1.3$\pm$0.5kms$^{-1}$} for LHS 3808 and -1.7$\pm$0.5kms$^{-1}$ for LHS 3809.

\section{Summary and Conclusions}
\label{sec:conclusion}
We constructed a sample of 67 wide, fully convective M-dwarf binaries using Gaia kinematics and measured H\textalpha\ equivalent widths for each component with FAST, a mid-resolution optical spectrograph. We classified pairs as active, inactive, or transition based on the equivalent widths of their H\textalpha\ features. We then followed up candidate transition systems using higher-resolution spectroscopy to vet them for unresolved binaries. Ultimately, we found 22 systems in which both components are active, 36 systems in which both are inactive, 8 transition systems with one active and one inactive component, and 1 newly discovered triple, LDS 942AC-B.

We gathered ten epochs of spectra for LDS 942AC with the $R=44000$ TRES spectrograph in order to fit the orbit of this new double-lined spectroscopic binary. We found that the pair orbits with a period of 25 days and is substantially eccentric. Intriguingly, the most massive star in the LDS 942 system is the widely separated B component, which is also highly H\textalpha\ active. The A component is less massive with modest H\textalpha\ emission, and the C component is least massive with no measurable H\textalpha\ emission. Many examples are known in which an M dwarf in a close binary has its H\textalpha\ emission and rapid rotation persist to advanced ages due to interactions between the binary components. This system suggests that an opposite effect may also be possible: while we typically expect the more massive component to spin down first, interactions between the A and C component may have caused them to spin down at a younger age.

Next, we presented an empirical relationship to estimate the average epoch of spindown for fully convective M dwarfs (Equation~\ref{eq:epoch}), which is the mass-dependent age of transition between the active/rapidly rotating mode and the inactive/quiescent one. This equation is based on a volume-complete sample of 323 single, 0.1--0.3M$_\odot$ M dwarfs that we studied in \citet{Pass2023a, Pass2023} and hence should only be applied to fully convective M dwarfs. Using this relationship, we analyzed our wide binary sample to determine if our observations were consistent with the epoch of spindown being dependent on stellar mass alone, specifically for these systems in which metallicity and birth environment are presumably the same for both components.

We observed more active pairs and fewer inactive pairs than we naively expected, and determined that biases due to youth or unresolved binarity were unlikely to explain the discrepancy. However, the observed distribution of systems was consistent with dynamical processing, with the population of wide binaries dwindling with age due to gravitational perturbations by passing stars causing some binaries to be disrupted. Accounting for this effect, we found that the fraction of transition pairs was consistent with the spindown epoch depending on stellar mass alone; however, the number of equal-mass binaries that were transition systems was marginally higher than expected, suggesting that some second-order effects may also be present. A stochastic component with a dispersion of $\leq$0.5Gyr (or $\leq$25\%) would be consistent with our population-level observations. While we refer to this component as stochastic, it is not necessarily random; rather, such a dispersion could result from deterministic but unknown differences in the early evolution of the stars, such as the influence of planet formation.

When controlling for metallicity and birth environment, the epoch of spindown for fully convective M dwarfs is therefore predominately determined by stellar mass. However, we know that for a given stellar mass, there is significant dispersion in the spindown epochs of field stars: in \citet{Pass2022}, we identified a sample of 0.3M$_\odot$ stars that had spun down by 600Myr, while Equation~\ref{eq:epoch} yields a typical spindown epoch of 1.3Gyr for this mass. Further work to quantify the dispersion in the field sample is necessary to determine whether the field's star-to-star variation exceeds the stochastic component allowed by our observations of M-M binary pairs. If the field sample has a larger dispersion, it would imply that the dominant source of variation is not stochastic, but rather based on factors that are shared between members of our wide binary pairs. For example, the high-energy birth environment may determine the circumstellar disk lifetime, and hence, the length of the disk-locking phase and the initial rotation of the star following disk dissipation \citep{Roquette2021}. Such a result could follow from magnetic-morphology-driven spindown \citep{Garraffo2018, Monsch2023}, in which the epoch of spindown is determined entirely by initial rotation rate and stellar mass, or the \citet{Matt2015} torque law, which similarly depends on these parameters. A significant environmental influence would also be consistent with observations of cluster-to-cluster variance in the distribution of stellar rotation rates \citep[e.g.,][]{Coker2016}. Given such cluster observations, it is likely reasonable to assume that initial rotation rate has both an environmental and individual component, with the high-energy birth environment shared between the two stars in our binaries playing a role, as well as differences in the early evolution of each star.

We reported the properties of our stars in Tables~\ref{tab:transition}, ~\ref{tab:active}, and \ref{tab:inactive}. Our eight transition systems are exciting targets for further study, representing pre- and post-spindown laboratories that are controlled for confounding variables such as differences in metallicity, birth environment, and age—and in the case of the two equal-mass binaries, also mass. To facilitate future work on these systems, we concluded by summarizing previous observations of these stars from the literature, as well as new inferences from TESS photometry and our TRES spectroscopy.

\section*{Acknowledgements}
We thank Allyson Bieryla, Warren Brown, Lars Buchhave, Pascal Fortin, Jonathan Irwin, Amber Medina, Sean Moran, Samuel Quinn, Andrew Szentgyorgyi, and Jennifer Winters for their contributions to FAST/TRES operations and/or code development.

This paper uses data products produced by the OIR Telescope Data Center, supported by the Smithsonian Astrophysical Observatory; the Two Micron All Sky Survey, which is a joint project of the University of Massachusetts and the Infrared Processing and Analysis Center/California Institute of Technology funded by the National Aeronautics and Space Administration and the National Science Foundation; the TESS mission, with funding provided by NASA's Science Mission Directorate and public data processed at the TESS Science Office and the TESS Science Processing Operations Center; and the European Space Agency (ESA) mission Gaia, processed by the Gaia Data Processing and Analysis Consortium (DPAC). Funding for the DPAC has been provided by national institutions, in particular the institutions participating in the Gaia Multilateral Agreement. EP is supported in part by a Natural Sciences and Engineering Research Council of Canada (NSERC) Postgraduate Scholarship. 

\facilities{FLWO:1.5m (FAST/TRES), Gaia}
\software{\texttt{BANYAN \textSigma} \citep{Gagne2018}, \texttt{exoplanet} \citep{ForemanMackey2021}, \texttt{galpy} \citep{Bovy2015}, \texttt{Lightkurve} \citep{lightkurve2018}, \texttt{Matplotlib} \citep{Hunter2007}, \texttt{NumPy} \citep{Harris2020}, \texttt{pandas} \citep{Reback2021}, \texttt{PyAstronomy} \citep{Czesla2019}, \texttt{PyMC3} \citep{Salvatier2016}, \texttt{SciPy} \citep{Scipy2020}}

\bibliography{mm}{}
\bibliographystyle{aa_url}



\appendix
\section{Long tables}
\restartappendixnumbering

\begin{deluxetable*}{rrrrrrrrrrr}[b]
\label{tab:transition}
\tabletypesize{\footnotesize}
\tablecolumns{11}
\tablewidth{0pt}
 \tablecaption{Transition systems}
 \tablehead{
 \colhead{ \vspace{-0.1cm}2MASS A} & 
 \colhead{ \vspace{-0.1cm}2MASS B} &
 \colhead{$d$} &
 \colhead{$\rho$} &
 \colhead{$M_A$} &
 \colhead{$M_B$} &  
 \colhead{$\Delta M$} &
 \colhead{FAST-H\textalpha$_{A}$} & 
 \colhead{FAST-H\textalpha$_{B}$} &
 \colhead{TRES-H\textalpha$_{A}$} & 
 \colhead{TRES-H\textalpha$_{B}$}
 \\
 \colhead{} &
 \colhead{} &
 \colhead{[pc]} &
 \colhead{["]} &
 \colhead{[M$_\odot$]} &
 \colhead{[M$_\odot$]} &
 \colhead{[M$_\odot$]} &
 \colhead{[\AA]} &
 \colhead{[\AA]} &
  \colhead{[\AA]} &
 \colhead{[\AA]}}
\startdata
\\
11231269+8009027&11231650+8009045&26&10&0.151&0.150&0.002&-0.695$\pm$0.058&-0.151$\pm$0.060&-1.691&-0.192 \\
10204884$-$0633195&10205111$-$0634400&31&87&0.256&0.240&0.016&-3.571$\pm$0.044&0.020$\pm$0.026&-3.746&0.064 \\
18524373+3659257&18524397+3659176&23&9&0.161&0.130&0.031&-0.041$\pm$0.034&-1.598$\pm$0.057&-0.196&-1.765 \\
01232866+6411443&01233090+6411440&32&15&0.312&0.270&0.042&0.106$\pm$0.028&-1.279$\pm$0.025&0.074&-1.687 \\
14593063+2833387&14593085+2833463&33&8&0.209&0.157&0.052&0.018$\pm$0.032&-2.712$\pm$0.054&0.049&-2.207 \\
20350677+0218289&20350608+0218166&20&16&0.295&0.188&0.107&0.074$\pm$0.041&-3.982$\pm$0.042&-0.014&-4.049 \\
16552527$-$0819207&16553529$-$0823401&6&299&0.211&0.091&0.120&0.173$\pm$0.017&-4.863$\pm$0.075&0.071&-6.605 \\
22261576+0300182&22261549+0300075&23&12&0.332&0.135&0.196&0.017$\pm$0.024&-4.669$\pm$0.123&0.084&-6.035 \\
\enddata
\vspace{0.2cm}
\end{deluxetable*}

\begin{deluxetable*}{rrrrrrrrrrr}[h]
\label{tab:active}
\tabletypesize{\footnotesize}
\tablecolumns{11}
\tablewidth{0pt}
 \tablecaption{Active systems}
 \tablehead{
 \colhead{ \vspace{-0.1cm}2MASS A} & 
 \colhead{ \vspace{-0.1cm}2MASS B} &
 \colhead{$d$} &
 \colhead{$\rho$} &
 \colhead{$M_A$} &
 \colhead{$M_B$} &  
 \colhead{$\Delta M$} &
 \colhead{FAST-H\textalpha$_{A}$} & 
 \colhead{FAST-H\textalpha$_{B}$} &
 \colhead{TRES-H\textalpha$_{A}$} & 
 \colhead{TRES-H\textalpha$_{B}$}
 \\
 \colhead{} &
 \colhead{} &
 \colhead{[pc]} &
 \colhead{["]} &
 \colhead{[M$_\odot$]} &
 \colhead{[M$_\odot$]} &
 \colhead{[M$_\odot$]} &
 \colhead{[\AA]} &
 \colhead{[\AA]} &
  \colhead{[\AA]} &
 \colhead{[\AA]}}
\startdata
\\
22353648+0032374&22353647+0032332&41&4&0.286&0.285&0.001&-4.925$\pm$0.061&-3.539$\pm$0.063&& \\
12212705+3038357&12212673+3038376&12&4&0.125&0.123&0.002&-5.109$\pm$0.040&-4.704$\pm$0.044&& \\
04331742+6846543&04331837+6846573&19&6&0.131&0.127&0.004&-3.656$\pm$0.060&-2.088$\pm$0.061&& \\
06141246$-$1436023&06141237$-$1436085&26&6&0.212&0.207&0.005&-2.889$\pm$0.058&-2.007$\pm$0.058&& \\
19522690+3155187&19522688+3158119&33&173&0.309&0.304&0.005&-3.811$\pm$0.052&-5.585$\pm$0.058&& \\
14022709+1520339&14022674+1520384&35&7&0.283&0.277&0.006&-4.808$\pm$0.062&-6.072$\pm$0.074&& \\
14275607$-$0022310&14275640$-$0022191&18&13&0.278&0.269&0.009&-9.067$\pm$0.057&-9.289$\pm$0.057&& \\
09191895+3831159&09191904+3831233&19&7&0.214&0.203&0.011&-5.407$\pm$0.032&-4.722$\pm$0.032&& \\
04283289+4157240&04283205+4157239&47&9&0.273&0.261&0.012&-14.713$\pm$0.071&-4.372$\pm$0.059&& \\ 
03595303+1325443&03595281+1325415&34&4&0.287&0.270&0.018&-3.924$\pm$0.053&-5.768$\pm$0.048&& \\ 
21523313+1147445&21523345+1147460&50&5&0.316&0.289&0.027&-5.913$\pm$0.065&-16.596$\pm$0.103&& \\ 
17274706+5200018&17274680+5200079&29&6&0.232&0.203&0.029&-6.827$\pm$0.049&-8.524$\pm$0.060&& \\
10241364+3902333&10241320+3902304&18&6&0.258&0.224&0.034&-2.958$\pm$0.029&-3.577$\pm$0.034&& \\
13360002+4024118&13355969+4021459&24&5&0.178&0.141&0.037&-4.924$\pm$0.033&-7.021$\pm$0.061&& \\
00095737$-$0636149&00095982$-$0632010&35&256&0.269&0.227&0.042&-4.582$\pm$0.051&-4.939$\pm$0.066&& \\
02090447+4341267&02090486+4341250&37&5&0.327&0.284&0.043&-5.703$\pm$0.061&-2.820$\pm$0.063&& \\
02230174+6600452&02230030+6600446&42&9&0.215&0.168&0.046&-6.089$\pm$0.079&-4.976$\pm$0.130&& \\
02032589+0648008&02033222+0648588&24&111&0.271&0.211&0.060&-0.785$\pm$0.025&-2.714$\pm$0.036&-1.465&-2.423 \\
12332604+5226589&12332593+5227167&46&18&0.308&0.239&0.069&-5.979$\pm$0.042&-5.644$\pm$0.073&& \\
23204335+8329513&23204037+8329463&46&7&0.289&0.220&0.069&-3.635$\pm$0.090&-5.139$\pm$0.094&& \\ 
09234719+6357447&09234740+6357387&32&6&0.278&0.164&0.115&-4.917$\pm$0.038&-3.977$\pm$0.045&& \\
03513447+0722250&03513420+0722229&36&5&0.298&0.169&0.129&-4.018$\pm$0.075&-6.169$\pm$0.142&& \\
\enddata
\vspace{0.2cm}
\end{deluxetable*}

\begin{deluxetable*}{rrrrrrrrrrr}[h]
\label{tab:inactive}
\tabletypesize{\footnotesize}
\tablecolumns{11}
\tablewidth{0pt}
 \tablecaption{Inactive systems}
 \tablehead{
 \colhead{ \vspace{-0.1cm}2MASS A} & 
 \colhead{ \vspace{-0.1cm}2MASS B} &
 \colhead{$d$} &
 \colhead{$\rho$} &
 \colhead{$M_A$} &
 \colhead{$M_B$} &  
 \colhead{$\Delta M$} &
 \colhead{FAST-H\textalpha$_{A}$} & 
 \colhead{FAST-H\textalpha$_{B}$} &
 \colhead{TRES-H\textalpha$_{A}$} & 
 \colhead{TRES-H\textalpha$_{B}$}
 \\
 \colhead{} &
 \colhead{} &
 \colhead{[pc]} &
 \colhead{["]} &
 \colhead{[M$_\odot$]} &
 \colhead{[M$_\odot$]} &
 \colhead{[M$_\odot$]} &
 \colhead{[\AA]} &
 \colhead{[\AA]} &
  \colhead{[\AA]} &
 \colhead{[\AA]}}
\startdata
\\
06492207+3209599&06492214+3209551&31&5&0.236&0.235&0.001&-0.008$\pm$0.038&0.050$\pm$0.041&& \\
19071320+2052372&19070556+2053168&9&114&0.346&0.345&0.001&0.037$\pm$0.010&0.045$\pm$0.010&& \\ 
20252724+5433327&20252445+5433458&32&28&0.316&0.312&0.005&0.077$\pm$0.038&0.114$\pm$0.038&& \\
22113258+0058490&22113228+0058514&34&5&0.265&0.258&0.007&0.189$\pm$0.029&0.127$\pm$0.030&& \\
01431772$-$0151223&01431760$-$0151265&47&5&0.227&0.214&0.013&-0.080$\pm$0.053&-0.894$\pm$0.061&-0.388&-0.390 \\
14452674$-$1311353&14452576$-$1311495&33&20&0.308&0.292&0.015&-0.003$\pm$0.033&0.128$\pm$0.035&& \\
10260265+5027090&10260331+5027220&19&14&0.301&0.277&0.023&0.026$\pm$0.042&0.010$\pm$0.046&& \\
12204478$-$0451408&12204459$-$0451456&34&6&0.238&0.213&0.024&0.060$\pm$0.036&0.036$\pm$0.043&& \\
19384867+3512361&19384898+3512370&34&4&0.168&0.144&0.024&0.752$\pm$0.044&0.507$\pm$0.091&& \\ 
14364366$-$0830245&14364401$-$0830305&44&8&0.330&0.305&0.025&0.032$\pm$0.052&0.113$\pm$0.059&& \\
02513445+5922325&02513375+5922349&36&6&0.263&0.236&0.026&-0.153$\pm$0.042&-0.205$\pm$0.053&& \\
13102287+3155167&13102342+3155159&50&7&0.322&0.292&0.030&-0.112$\pm$0.030&0.118$\pm$0.044&& \\
03263459+3929072&03263418+3929029&19&7&0.242&0.210&0.032&0.362$\pm$0.025&0.252$\pm$0.030&& \\
02541919+6427401&02542325+6425552&43&108&0.312&0.277&0.035&0.023$\pm$0.047&-0.173$\pm$0.100&& \\
18251482+0721284&18250818+0721482&37&101&0.341&0.306&0.036&0.201$\pm$0.031&0.335$\pm$0.035&& \\
14022402$-$0312001&14022282$-$0312217&32&28&0.198&0.159&0.039&0.038$\pm$0.050&0.141$\pm$0.057&& \\
21334913+0146561&21334914+0147012&17&5&0.213&0.173&0.040&0.082$\pm$0.027&0.083$\pm$0.029&& \\
23580162+7836301&23575233+7836458&26&32&0.307&0.266&0.041&0.070$\pm$0.031&0.199$\pm$0.038&& \\
02330114+0105389&02330063+0106070&45&29&0.257&0.211&0.046&0.039$\pm$0.057&-0.170$\pm$0.078&& \\
18115228+3225199&18115554+3225466&28&49&0.194&0.145&0.049&0.063$\pm$0.058&-0.012$\pm$0.054&& \\
20563492+3047518&20563466+3047425&33&10&0.293&0.241&0.052&0.073$\pm$0.028&0.074$\pm$0.033&& \\
10230158$-$0735092&10230164$-$0735248&37&16&0.235&0.183&0.053&-0.075$\pm$0.044&-0.081$\pm$0.053&& \\
00515271+3750216&00515693+3751159&38&74&0.279&0.225&0.054&0.056$\pm$0.052&-0.137$\pm$0.068&& \\
22585768+6430048&22573941+6418533&48&841&0.332&0.255&0.078&0.083$\pm$0.039&0.257$\pm$0.052&& \\
22282013+0303534&22281858+0303424&49&26&0.305&0.224&0.081&0.196$\pm$0.043&0.432$\pm$0.063&& \\
10191279$-$0305520&10191244$-$0305519&46&5&0.252&0.165&0.087&0.139$\pm$0.048&-0.017$\pm$0.040&& \\
15400352+4329396&15400374+4329355&13&4&0.295&0.204&0.091&0.334$\pm$0.012&0.134$\pm$0.032&& \\
21440900+1703348&21440795+1704372&17&64&0.264&0.171&0.093&0.154$\pm$0.037&-0.079$\pm$0.059&& \\
12114753+2400064&12114707+2400054&40&7&0.253&0.158&0.096&0.296$\pm$0.034&0.238$\pm$0.080&& \\
11485296+1800581&11485323+1800564&42&4&0.347&0.249&0.098&0.143$\pm$0.052&-0.078$\pm$0.047&& \\
04281667+0600130&04281665+0600176&24&5&0.260&0.130&0.130&0.071$\pm$0.040&-0.198$\pm$0.084&& \\
02274112+0613539&02274066+0613550&40&7&0.330&0.189&0.141&0.047$\pm$0.061&-0.146$\pm$0.095&& \\
14515365+5147107&14515297+5147134&33&7&0.284&0.136&0.148&0.313$\pm$0.031&0.370$\pm$0.050&& \\
18180427+3846342&18180345+3846359&11&10&0.318&0.165&0.152&0.165$\pm$0.019&0.093$\pm$0.022&& \\
15352059+1742470&15352039+1743045&15&18&0.297&0.129&0.168&0.117$\pm$0.016&0.036$\pm$0.038&& \\
19445022$-$0203562&19444998$-$0203534&37&5&0.345&0.161&0.184&0.315$\pm$0.044&0.133$\pm$0.059&& \\
\enddata
\vspace{1cm}
\end{deluxetable*}

\end{document}